\documentclass[12pt,emtex]{article}
\usepackage{amsfonts}
\usepackage{amssymb}
\usepackage{amscd}
\newcommand{\BEQ}{\begin{equation}}
\newcommand{\EEQ}{\end{equation}}

\newtheorem{Th}{Theorem}
\newtheorem{lem}{Lemma}

\newtheorem{Ex}{Example}

\newtheorem{Rem}{Remark}

\def\bea{\begin{eqnarray}}
\def\eea{\end{eqnarray}}

\def\S{{\Sigma}}
\def\O{{\cal O}}
\def\C{{\mathbb{ C}}}

\def\Z{\mathbb{Z}}

\def\L{\mathcal{L}}
\def\l{\lambda}
\def\m{\mu}
\def\K{\mathcal{K}}

\def\M{\mathcal{M}}
\def\T{\mathcal{T}}

\begin{document}

\begin{titlepage}
\hfill ITEP-TH-22/03

\vskip 3.0cm

\centerline{\Large \bf Riemann bilinear form and Poisson structure}
\vskip 0.4cm \centerline{\Large \bf in Hitchin-type systems.}
\vskip 1.0cm \centerline{D. Talalaev \footnote{E-mail:
talalaev@gate.itep.ru} }
\vskip 0.2cm
\centerline{\sf Institute for Theoretical and Experimental Physics
\footnote{ITEP, 25 B. Cheremushkinskaya, Moscow, 117259, Russia.}}
\vskip 2.0cm
{\bf Abstract.} In this paper we reinterpret the Poisson
structure of the Hitchin-type system in cohomological terms.
The principal ingredient of a new interpretation in the case of the Beauville system
is the meromorphic cohomology
of the spectral curve, and the main result is the identification of the
Riemann bilinear form and the symplectic structure of the model.
Eventual perspectives of this approach lie in the quantization domain.

\vskip 1.0cm \tableofcontents

\end{titlepage}

\section*{Introduction}
For the moment there are too many different examples of integrable systems.
The methods of resolution whenever exist bear no resemblance. However the
so-called Hitchin-type description fits into such systems as the Toda
chain, the Neumann system, all versions of the spin Calogero-Moser system
\cite{BBKT},
the rational and
elliptic Gaudin model with spin \cite{T1}.
This observation encourage us to develop this language and use this
type of coordinates in the quantization problem due to its relative generality.

This paper is organized as follows. In the first section we describe two
types of considered models -- the Hitchin system and the dynamical system
on the space of rational matrices. We recall the construction of
separated variables and ``action-angle'' variables. In this section we
represent the Calogero-Moser system in terms of rational matrices. The
second section is devoted to the symplectic structure of the Hichin system. We
construct such a factorization of the tangent space to the phase space of
the system which identify the symplectic form with the Serre pairing on the
spectral curve between $H^0(\K_\S)$ and $H^1(\O_\S).$ In the third section
we give the analogous but more explicit factorization of the tangent space
to the phase space of the integrable system on the space of rational
matrices. In this case we also relate the tangent space with the
meromorphic cohomology group of the spectral curve and show that the
symplectic structure coincides with the Riemann bilinear form.
The conclusion contains some open
questions in the problem of quantization.
\vskip 1cm
{\bf Acknowledgments.} The author is grateful to O. Babelon and F. Smirnov
for fruitful discussions along the work on this paper; to I. Krichever for
numerous advises, especially for the example of the lagrangian connection; to
A. Chervov, A. Gorodentsev, A. Samohin and S. Loktev for useful remarks.
A part of this work was done during the author's staying at LPTHE
(Universit\'e Paris 6).
This work was partly supported by RFBR grant 01-01-00546, RFBR grant for
support of young scientist 03-01-06236.

\section{Model description}
\subsection{Hitchin-type systems}
Originally the Hitchin system was defined in \cite{Hit} as an integrable
system on the
cotangent bundle of the moduli space of stable holomorphic bundles of fixed
rank $r$ and degree $d$ over the algebraic curve $\S_0.$
The Poisson structure is the canonical one on the cotangent bundle.
The space of function in involution is constructed
as follows: a cotangent vector at a point $E$ of the moduli space
$\mathcal{M}$ is an element $\phi\in H^0(End(E)\otimes\mathcal{K}).$
There is a well defined function $h_i:\mathcal{T}^*\mathcal{M}\rightarrow
H^0(\mathcal{K}^{\otimes k})$ such that $h_i(E,\phi)=\frac 1 i tr\phi^i.$ The map
$$h:\mathcal{T}^*\mathcal{M}\longrightarrow\oplus_{i=1}^r
H^0(\mathcal{K}^{\otimes i})$$
is called Hitchin map and realize the algebraic integrability, which means
that the fibers are abelian lagrangian varieties of half dimension.
The crucial role is played by the spectral curve that is defined as follows.
One can define the bundle morphism (non-linear) $char(\phi):\mathcal{K}
\rightarrow\mathcal{K}^{\otimes r}$ by
$$char(\phi)(\mu)=det(\phi-\mu* Id)$$
where $\mu$ is a local section of $\mathcal{K}$ and $Id$ is the
identical global section of $End(E).$ The spectral curve is the
preimage of the zero section in $\mathcal{K}^{\otimes r}.$ It is an
algebraic curve $\S$ in the projectivization of the total space of $\mathcal{K}.$

The main problem in this description is the explicit parameterization of the
moduli space. There are several achievements in this direction. Firstly one
can parameterize holomorphic bundles by \v{C}ech cochains. A problematic
step is an infinite dimensional reduction to cohomologies.
Another approach works in some specific situations \cite{GP,GAW} and
relates to the Narasimhan-Ramanan parameterization \cite{NR1,NR2}. The
Hecke-Turin coordinates \cite{Kr,ER} provide another geometrical way to parameterize the
moduli space.

The first level generalization of the Hitchin construction occurs when the
base curve is singular.
This remarkable situation provides the
explicit parameterization of the moduli space
in purely linear-algebraic terms. By the other hand it is rather general:
such systems as the rational, trigonometric, and elliptic Calogero-Moser
models,the
rational and elliptic Gaudin systems, the Neumann system and the
Toda chain arise in the framework of the Hitchin description on singular curves.
This approach appeared in
\cite{NN} and was elaborated in \cite{CT,CT1}.

The second level generalization crops up in the
context of the space of Higgs pairs. A Higgs pair consists of
a holomorphic bundle $E$ on
$\S_0$ and a bundles homomorphism $\phi:E\longrightarrow E\otimes L,$ where
$L$ is some fixed line bundle on $\S_0$ not necessarily the canonical class. The
Poisson structure is less geometric in this case, it is given by the
Kostant-Kirillov
brackets on orbits of the loop group \cite{Bea,DM,HH}. We show that
both generalizations coincide in several examples and it is the subject of
future works to establish their general correspondence.

\subsection{Rational matrices and separated variables}
In \cite{BT1} the following description of integrable
systems was proposed: let us consider a family $Spec$ of
spectral curves $\S_H$ defined by the equation
\begin{equation}\label{cur}
R(\l,\m)=R_0(\l,\m)+\sum_{i=1}^g H_i R_i(\l,\m)=0
\end{equation}
where $H_i$ are parameters and $R_i$ are fixed polynomials on two variables.
An analog of the Hitchin fibration can be obtained in these terms: one
takes the space $\C^{2g}$ which is the space of $g$-tuples of points $P_i$
in $\C^2$ with coordinates $(\l_i,\m_i).$ A condition that a curve $\S\in
Spec$ contains $g$ points $P_i$ determines this curve $$H=-B^{-1}V$$ where
$B_{ij}=R_j(\l_i,\m_i)$ and $V_i=R_0(\l_i,\m_i).$
Additionally, $g$ points $P_i$ gives us a divisor $D$ of degree $g$ on this
curve. Hence one has the following fibration
$$
\begin{CD}
\C^{2g}\\ @VV{Jac(\S)}V\\
Spec
\end{CD}
$$
The total space is endowed with a family of Poisson structures
$\{\l_i,\m_j\}=\delta_{ij}f(\l_i,\m_i).$ One of the propositions of
\cite{BT1} is that this construction provides an integrable system for all
choices of such Poisson brackets, i.e. that the quantities $H_i$ are in
involution. In this case the problem of constructing ``angle'' variables
can be solved by the following
{\lem The quantities $$\phi_i=\sum_{k}\int^{\l_k}\frac {R_i(\l,\m)d\l}
{\partial_{\m}R(\l,\m)f(\l,\m)}$$ are conjugated to $H_i$
$$\{H_m,\phi_l\}=\delta_{ml}.$$}
\\
{\bf Proof.} Indeed, the equation (\ref{cur}) at the point $P_j$ imply
$$R_0(\l_j,\m_j)+\sum_i H_i R_i(\l_j,\m_j)=0$$ and
$$\partial_{\m}R(\l_j,\m_j)\{\l_k,\m_j\}+\sum_i
R_i(\l_j,\m_j)\{\l_k,H_i\}=0.$$ Introducing the notation
$$C_{ik}=\{\l_k,H_i\},\quad\widetilde{B}_{ji}=-\frac {R_i(\l_j,\m_j)}
{\partial_{\m}R(\l_j,\m_j)f(\l_j,\m_j)}$$ one obtains equivalently
$$\widetilde{B} C=I$$ where $I$ is the identity matrix. Now we calculate
the Poisson bracket $$\{H_m,\phi_l\}=-\sum_i \{\l_i,H_m\}
\frac{R_l(\l_i,\m_i)}{\partial_{\m}R(\l_i,\m_i)f(\l_i,\m_i)}=
\sum_i C_{mi}\widetilde{B}_{il}=\delta_{ml}$$ due to uniqueness of an
inverse matrix.$\square$

Let us show that such kind of description in terms of separated
variables arises from a generalization of the Hitchin system -- the
dynamical system on the space of Higgs pairs. We consider a specific case
of the so-called Beauville system. Let us
consider the rational curve $\C P^1$ and the degree
$n$ linear bundle  $\O(n)$ on it. The semistable bundles on $\C P^1$ can be
represented in the form $E=\O^r\otimes\O(m)$ where $\O^r$ is the trivial
holomorphic bundle of rank $r.$ A Higgs field is an element of $H^0(\O^r\otimes
\O(n))$ due to the fact that $End(E)$ is trivial. On the space of such
sections there is a family of Poisson structures given as follows: for
a divisor $D$ of degree $n$ there is a realization of the space of
holomorphic section $H^0(\O^r\otimes\O(n))$ by rational matrices with poles
at $D=\sum_{i=1}^k n_i z_i.$ This can be done by taking a trivialization
of the bundle $\O^r\otimes \O(n)$ over the affine part $\C P^1/D.$
On the space of rational matrices with prescribed poles
$$\Phi(z)=\sum_{i=1}^k\Phi_i(z)+\Phi(\infty);
\qquad \Phi_i(z)=\sum_{j=1}^{n_i}\frac{\Phi_i^{j}}{(z-z_i)^{j}},$$
there is a Poisson structure given by the
Kostant-Kirillov form on co-adjoint orbits of the loop group represented
by polar parts $\Phi_i(z).$ The involutive subalgebra of functions is spanned
by coefficients of the characteristic polynomial. The casimirs
of co-adjoint orbits are among them.
There are too many integrals and that is why one fixes some of them to be central.
Different choices of a divisor and central elements imply different
integrable systems. Let us demonstrate it on examples.

{\Ex~}Consider a divisor $D=z_1+z_2+z_3-\infty$ of degree $2.$ The
corresponding Higgs field is
$$\Phi(z)=\frac {\Phi_1}{z-z_1}+\frac {\Phi_2}{z-z_2}+\frac{\Phi_3}{z-z_3}$$
with the condition $\Phi_1+\Phi_2+\Phi_3=0.$ Let us choose central
elements to be \mbox{$c_l=Tr(\Phi_1^l)+Tr(\Phi_2^l)$} and fix its value $c_l=0.$
These conditions mean that $\Phi_1$ and $-\Phi_2$ belong to
the same orbit. In other
words there exists such $\Lambda$ that $\Phi_1\Lambda=-\Lambda\Phi_2.$
Resolving the condition $\Phi_1-\Lambda^{-1}\Phi_1\Lambda+\Phi_3=0$ one
obtains the traditional parameterization of the trigonometric Calogero-Moser
system (see \cite{CT} for details).

One can easily see that the spectral
curve $\S$ of the model $$det(\Phi(z)-k)=0,$$ which is constructed as a covering
of the rational curve $\C P^1,$ can be considered as a covering of a
singular curve with the double point. Indeed, the points of $\S$ over $z_1$ and
$z_2$ coincide, this implies that $k$ is a function on $\S_{sing},$ where
the singularities on $\S_{sing}$ arise from the double point on $\C P^1.$
The Higgs field in this case corresponds to an element of
$H^0(End(E)\otimes\K)$ on the rational curve with the double point for some
holomorphic bundle $E$ over it given by $\Lambda.$

{\Ex~}Consider a divisor $D=2z_1+z_2-\infty$ of degree $2.$ The
corresponding Higgs field is
$$\Phi(z)=\frac {\Phi_0}{(z-z_1)^2}+\frac {\Phi_1}{z-z_1}+\frac{\Phi_2}{z-z_2}$$
with the condition $\Phi_1+\Phi_2=0.$ Let us choose central
elements to be $c_l=Tr(\Phi_0^l\Phi_1)$ and fix its values $c_l=0.$ It means
that the orbit of $\Phi_1(z)$ contains an element of the
form $$\frac {A_0}{(z-z_1)^2}.$$ In other
words there exists such $X$ that $\Phi_1=[X,\Phi_0].$
Resolving the condition \mbox{$\Phi_1+\Phi_2=0$} one
obtains the traditional parameterization of the rational Calogero-Moser
system (see \cite{CT} for details).

By analogy with the trigonometric
case the spectral
curve $\S$ of the model, which is constructed as a covering
of the rational curve $\C P^1,$ can be projected to a covering of a
singular curve with the cusp singularity. The Higgs field in this case
can be realized as an element of
$H^0(End(E)\otimes\K)$ on the rational curve with the cusp singularity for some
holomorphic bundle $E$ over it given by $X$. Coincidence of the
canonical  form on co-adjoint orbits and the form in
terms of separated variables was proved in \cite{BT2}.

\section{Canonical symplectic form. Hitchin system}
\subsection{Factorization}
We start with an identification between deformations of
the spectral curve and holomorphic differentials on it. For a general
algebraic curve this relationship can not be established,
one has $(3g-3)$-dimensional
space of deformations and only $g$-dimensional space
of holomorphic differentials. The spectral curves of considered
systems are special; they are parameterized by a $g$-dimensional
space. One has the Hitchin map
$$h:\mathcal{T}^*\mathcal{M}\longrightarrow
\bigoplus_{i=1}^{r}H^0(\S_0,\mathcal{K}^{\otimes i})$$ where the right
hand side corresponds to the space of spectral curves $Spec.$ It is a linear space
and one can identify its tangent space $\mathcal{T}Spec$ with itself.
One also has that the direct image of the structure sheaf $\O_{\S}$ of the
spectral curve is
\begin{equation}\label{ide2}
\pi_*(\O_{\S})=\O_0\oplus\mathcal{K}_0^{-1}\oplus\ldots\oplus\mathcal{K}_0^{1-r}
\end{equation}
where we have denoted $\O_0=\O_{\S_0}$ and $\mathcal{K}_0=\mathcal{K}_{\S_0}$ and
$\S_0$ is the base curve. This expresses the fact that locally any function
on $\S$ is a polynomial of degree $<r$ on a fiber of the canonical class
with coefficients which are functions on the base curve $\S_0.$
This consideration provides the following isomorphisms
$$H^0(\O_\S)\cong H^0(\pi_*(\O_\S));\qquad H^1(\O_\S)\cong H^1(\pi_*(\O_\S))$$
due to the fact that the spectral sequence $H^i(\S_0,R^j \pi_*\mathcal{F})$
converges on the first term for coverings of curves.
By virtue of Serre duality the second isomorphism is equivalent to the
following
\begin{equation}\label{ide1}
H^0(\mathcal{K}_\S)\cong H^0(\pi_*(\O_\S)^*\otimes \mathcal{K}_0)=
H^0(\oplus_{i=1}^r\mathcal{K}_0^{\otimes i})=\mathcal{T}Spec.
\end{equation}
The differential of the Hitchin map $dh:\T_{(E,\Phi)}(\T^*\M)
\longrightarrow H^0(\K_{\S})$ has an explicit description occurred in the
context of the Whitham hierarchy in \cite{Kr1}.
As it was mentioned in the introduction the spectral curve $\S$ lies in the
compactification $\widetilde {\K_0}$ of the total space of the canonical class
of the base curve
$\S_0.$

One has the tautological section $y\in H^0(\widetilde{\K_0},\pi^*(\K_0)).$
Its restriction to the spectral curve gives us a holomorphic differential
over $\S.$ The infinitesimal form of the Hitchin map becomes more clear in
a local picture. Let $U$ be an open set in $\S_0$ and $z$ -- a local
parameter on $U$. The canonical class trivializes over $U$ and the Higgs field
$\Phi\in H^0(End(E)\otimes \K_0)$ can be represented as $\Phi(z)=L(z)dz$
where $L(z)$ is a matrix valued function on $z.$ The spectral curve over $U$
is a set of pairs $(z,k)$ which solve the characteristic equation
$det(L(z)-k)=0.$ The tautological section in these terms is $y=kdz,$ which is
a well defined holomorphic differential over the spectral curve. Now deforming
$\Phi$ one obtains a deformation $\delta y:=\delta k dz$ where by
$\delta k$ we denote a variation of eigenvalues of $L.$ Furthermore one has the
following
{\lem \label{dif} Let $\delta \Phi\in\T_{(E,\Phi)}(\T^*\M)$ be a tangent vertical vector
deforming $\Phi$ then $$dh(\delta\Phi)=\delta y$$ under the
identification (\ref{ide1})
}
\\
{\bf Proof.} Let us choose such a covering of the base
curve $\S_0$ with two charts $U_1,~U_2$  that the intersection
$U=U_1\cap U_2$ does not contain brunching points. We choose also a
local parameter $z$ on $U.$ The canonical class $\K_0$ trivializes over
$U$ and we can choose a parameter $k$ along fibers. The representation
(\ref{ide2}) means that $\O_{\S}(\pi^{-1}U)$ consists of functions $f$ of
the type $$f=\sum_{i=1}^r f_i(z)k^{i-1};\quad
(\ref{ide2}):f\mapsto\tilde{f}=\{f_i|i=1,\ldots,r\}.$$
We will prove that $$<dh(\delta\Phi),\tilde f>=<\delta y, f>,$$
where on both sides we use appropriate Serre pairings.
$$<\delta y, f>=\sum_{k=1}^r\int_{\partial U_k} f \delta k dz=
\sum_{k=1}^r\int_{\partial U_k}\sum_{i=1}^r f_i(z) k^{i-1}\delta k dz,$$
where $U_k$ are
distinct preimages of $U:~\pi^{-1}U=\sqcup_k U_k.$ Using the fact that
$k$ takes the eigenvalues of $L$ over $U$ one obtains
$$<\delta y, f>=\sum_{k=1}^r\int_{\partial U_k}\sum_{i=1}^r  f_i(z)\frac 1 i \delta k^i
dz=\int_{\partial U} \sum_{i=1}^r f_i(z) \delta (\frac 1 i Tr L^i) dz.$$
Let us note that the expression on the right hand side is indeed the pairing
between $H^1(\O_0\oplus\K_0^{-1}\oplus\dots\oplus\K_0^{1-r})$ and
$\mathcal{T}Spec=H^0(\oplus_{i=1}^r\mathcal{K}_0^{\otimes i})~\square$
\vskip 0.5cm

Now we return to the Poisson description of the Hitchin system. Hamiltonian
flows correspond to the
dynamics of a line
bundle $\L$ over $\S$ constructed from the following exact sequence
$$0\longrightarrow\L(-R)\longrightarrow\pi^*E
\stackrel{\pi^*\Phi-y}\longrightarrow\pi^*(E\otimes\K_0)\longrightarrow 0,$$
where $R$ is the branching divisor of $\S$ over $\S_0.$
The line bundle $\L$ varies over $$Jac_{\S}=\frac
{H^1(\O_{\S})}{H^1(\S,\Z)}.$$
Hence the phase space can be fibered over $Spec$ by $Jac$ where over a
point $\S\in Spec$ the fiber is $Jac_\S.$
A tangent vector to $Jac_{\S}$ is an element of $H^1(\O_{\S}).$  Let us combine
the natural imbedding of vectors tangent to fibers to the tangent space of $\T^*\M$
at the point $(E,\Phi)$ and the differential of the Hitchin map
to the exact sequence of vector spaces:
\begin{equation}\label{ex1}
0\longrightarrow H^1(\O_{\S})\stackrel{\rho}\longrightarrow\T_{(E,\Phi)}(\T^*\M)
\stackrel{dh}\longrightarrow H^0(\K_{\S})\longrightarrow 0.
\end{equation}
{\lem \label{lem_ex} The exact sequence (\ref{ex1}) respects the special concordance
between the canonical symplectic
form $\omega$ on $\T_{(E,\Phi)}(\T^*\M)$ and the Serre pairing
$$<\cdot,\cdot>:~H^1(\O_{\S}) \times H^0(\K_{\S}) \rightarrow \C.$$
 Namely, for an element $\xi \in H^1(\O_{\S})$ and
$X\in\T_{(E,\Phi)}(\T^*\M)$ one has:
$$\omega(\rho(\xi),X)=<\xi,dh(X)>.$$
}
\\
{\bf Proof.} We consider the same
covering as in the previous lemma. Let us call ``vertical''
a tangent vector $v\in\T_{(E,\Phi)}(\T^*\M)$ which projects to $0$
on $\M$ by the differential of \mbox{$\tau:\T^*\M\longrightarrow\M.$}
One can see that each element of $\T_{(E,\Phi)}(\T^*\M)$ can be decomposed
into the sum of a vertical vector field and a hamiltonian vector field at a
point $(E,\Phi).$

The statement is obvious when $X$ is a value of a hamiltonian vector field.
It is equivalent to the fact that
the image $\rho(H^1(\O_{\S}))$ is
lagrangian. Indeed, there is a $g$-dimensional space of hamiltonian vector
fields which span the tangent space to the Jacobian at each point.
Such fields are orthogonal because the hamiltonians are in involution.

Let $X$ be a vertical tangent vector
$\delta \Phi=\delta L dz\in\T_{(E,\Phi)}(\T^*\M)$
and $\delta \L$ be a cocycle deforming
the line bundle on $\S.$ The direct image gives us a
deformation $\delta E=\pi_*\delta \L$ of a holomorphic bundle $E$ on the
base curve $\S_0.$ Let us calculate the symplectic form:
$$\omega(\rho(\delta \L),\delta \Phi)=\int_{\partial U}
Tr (\delta E\delta L) dz
=\sum_{i=1}^r\int_{\partial U_i}\delta\L\delta k dz=
<\delta \L,dh(\delta \Phi)>,$$
where the last equality is due to lemma \ref{dif}.
$\square$
{\Rem We present here the interpretation of the previous lemma
in terms of quasi-isomorphisms. Let us consider the diagram
$$
\begin{array}{ccccc}
&& H^1(\O_\S)\oplus\T_{(E,\Phi)}(\T^*\M) && \\
&\stackrel{p_1}\swarrow&& \stackrel{p_2}\searrow & \\
\T_{(E,\Phi)}(\T^*\M) &&&& H^1(\O_\S)\oplus H^0(\K_\S)
\end{array}
$$
where $p_1$ is a factorization subject to the diagonal embedding
$$0\longrightarrow H^1(\O_\S)\stackrel{(id\oplus\rho)\circ\Delta}
\longrightarrow
H^1(\O_\S)\oplus\T_{(E,\Phi)}(\T^*\M)\longrightarrow\ldots$$
and $p_2=id\oplus dh.$ Let $\omega'$ be the canonical symplectic form on
$H^1(\O_\S)\oplus H^0(\K_\S)$ defined by Serre duality. Then the lemma
above says that $$p_1^*\omega=p_2^*\omega'.$$
}
\subsection{Splitting}
One of principal questions in the context of the decomposition (\ref{ex1})
is a construction of a symplectic splitting of the exact sequence.
This means an embedding
$j:H^0(\K_\S)\longrightarrow\T_{E,\Phi}(\T^*\M)$ such that
$dh\circ j=id|_{H^0(\K_\S)}$
and
$$\rho\oplus j:H^1(\O_\S)\oplus H^0(\K_\S)\longrightarrow
\T_{E,\Phi}(\T^*\M)$$ is a symplectomorphism, where on the left hand side
the symplectic form is defined by Serre duality.
{\lem The space of such splittings is an open subset in the space of
lagrangian subspaces in $\T_{E,\Phi}(\T^*\M).$}
\\
{\bf Proof.} Let us note that the image $Im j$ is a lagrangian subspace.
Inversely, let $V$ be a lagrangian subspace in $\T_{E,\Phi}(\T^*\M)$ such
that $V\cap \rho(H^1(\O_\S))=0$ (it is an open condition). Then, choosing
a basis $\{e_1,\ldots,e_g\}$ in $H^0(\K_\S)$ and the dual basis
$\{e^*_1,\ldots,e^*_g\}$ in $H^1(\O_\S)$ let us construct the splitting $j$
as follows: $$j(e_i)=f_i$$ where $\{f_1,\ldots,f_g\}$
is the basis in $V\subset\T_{E,\Phi}(\T^*\M)$ dual to
$\{\rho(e^*_1),\ldots,\rho(e^*_g)\}.$ The condition $dh\circ
j=id|_{H^0(\K_\S)}$ fulfills due to lemma \ref{lem_ex}. $\square$
{\Rem In the rest of the paper we will consider several natural ways to
construct such a splitting. We also hope that the main ingredient
-- ``lagrangian connection'' -- plays a crucial role in the problem
of quantization.
}

\section{Canonical symplectic form. Rational matrices}
Firstly let us show that there is a natural lagrangian
connection in the case of Beauville system. The phase space is the space of
$g$ points $(\l_i,\m_i)$ in $\C^2.$ There is an exact sequence similar to the case
of the Hitchin system:
\begin{equation}\label{ex2}
\T Jac_\S\longrightarrow\T \C^{2g}\stackrel{dh}\longrightarrow\T_\S Spec.
\end{equation}
The first injection is lagrangian. A splitting
is an injection $j$ inverting $dh$:$$j:\T_\S Spec\hookrightarrow\T
\C^{2g}$$ such that $dh\circ j=id|_{\T_\S Spec}$ and $Im\,j$ is lagrangian. Let
us fix $\{\l_i\}$ and obtain a deformation of the spectral curve
and the divisor on it by varying only the $\m$-coordinates of points. This
deformation is obviously lagrangian due to the description of the Poisson
structure in terms of separated variables. This allows us to use the
following identification
$$\T \C^{2g}\cong H^1(\O_\S)\oplus H^0(\K_\S).$$
In this section we also identify tangent vectors to the
Jacobian with meromorphic differentials on the spectral curve by means of
the Baker-Akhiezer function and construct the isomorphism
\begin{equation}\label{mercoh}
\T \C^{2g}\cong H^0(\S,\Omega_{\frak{M}}^1)
\end{equation}
where $\Omega_{\frak{M}}^1$ is the sheaf of meromorphic $1$-forms
of the second kind on the
spectral curve. We prove that the symplectic
structure coincides with the Riemann bilinear form
in terms of the right hand side of (\ref{mercoh}).

\subsection{Linear dynamics on $Jac$ and Baker-Akhiezer function}
{\Rem Here we demonstrate the correspondence between the
Jacobian which is isomorphic to $Pic_0$ and the \v{C}ech
realization of the group of holomorphic degree $0$ line bundles in
$H^1(\S,\O^*).$ We have to construct a representative
cocycle for the point of $Jac.$ The main
technique is the Abel transform and its inverse.
}
\vskip 0.3cm
For the first let us consider the situation of
the dynamics linearized on the Jacobian by
components of the standard Abel transform. It means that
the coefficients $R_i$ define holomorphic differentials
on the spectral curve $\S.$ The dynamics on $Jac$ can
be represented by $X(t)=X(0)+tV.$ The vector $V$ corresponds to
the line bundle $L_0$ on $\S$ of degree $0$ constructed as follows.
The line bundles on $\S$ can be obtained by the inverse image of the
Abel transform $A^*$ from the line bundles on $Jac.$ The expression
$$s(X)=\frac {\theta(X+V)}{\theta(X)} $$ defines a line bundle on the
Jacobian and has monodromy properties
$$s(X+l)=s(X), \qquad s(X+\mathcal{B}l)=s(X)\exp(-2\pi i(l,V)), $$
where $l\in\Z^g$ and $\mathcal{B}$ is the matrix of $b$-periods of
holomorphic differentials. Now we fix a point $P_0$ on the
spectral curve and consider the space of normalized
meromorphic differentials
of the second kind with poles at $P_0.$ Its factor by exact
differentials is
$g$-dimensional. We can find such a
differential $\Omega_0$ which has specific $b$-periods
\begin{equation}
\label{mer}
\int_{b_i}\Omega_0=V_i.
\end{equation}
Now the expression
$$\psi(P)=s(A(P))\exp(2\pi i\int^P\Omega_0)$$ is a
well defined function on the spectral curve which
has an essential singularity at $P_0$ and
apart from $P_0$  its divisor is the Abel inverse of
$V.$ This function defines a line bundle on the spectral curve
with the same divisor as follows: we define a
covering $(U_0,U_1)$ of $\S$ such that $U_1=\S/P_0$ and $U_0$ is a disk centered
at $P_0$ which is sufficiently small to do not contain points of the
divisor. We take the principal part $\psi_{ess}$ of the function
$\psi$ at the point $P_0$ and declare it to be a transition
function for the chosen covering.
There is a meromorphic section $\psi$ on $U_1$ and $\psi/\psi_{ess}$ on
$U_0.$ It has the same divisor. We arrive to the classical lemma
(see \cite{S} and references therein)
{\lem The linear dynamics on the Jacobian can be represented in
terms of the transition function $f_{01}$ subject to the chosen
covering $(U_0,U_1)$ by follows
\begin{equation}
f_{01}(t)=f_{01}(0)\exp(t*ln(\psi_{ess})).
\end{equation}
}
\\
This lemma finishes the construction of the correspondence between the
tangent space of the Jacobian and meromorphic differentials of the second
kind with poles at $P_0.$ The choice of a family of spectral curves and a
divisor of poles of meromorphic differentials characterizes different types
of integrable systems. For example in the case of the Calogero-Moser system
one needs to take meromorphic differentials with poles of the second order
at points $P_i$ on the spectral curve which lie over $\infty\in\C P^1.$ But
in all cases there is the correspondence obtained above in terms of
cohomologies.

\subsection{Deformations of the spectral curve}
Let us consider the special case  of a spectral curve
$$R(\lambda,\mu)=R_0(\lambda,\mu)+\sum_i H_i R_i(\lambda,\mu)=0$$
where the differentials
$$\tilde\omega_i=\frac {R_i(\lambda,\mu) d\lambda}{\partial_\mu R(\lambda,\mu)}$$
are holomorphic.
{\Rem Let us note that all mentioned systems can be represented
in this form, at least birationally. An interesting question is
geometric interpretation of the dynamics when the differentials
linearizing hamiltonian flows are meromorphic. The case of the spin
generalization of the Calogero-Moser system gives us a partial response to
this question. In fact one should consider the generalized Jacobian where
the lattice is generated by the periods of holomorphic and
some meromorphic differentials.
And it can be identified with the Jacobian of a singular spectral curve
obtained by singularization of the base rational curve.
}

In this case the identification between differentials and tangent vectors
can be realized in a more direct way. The space of spectral curves is
parameterized by coordinates $H_i,$ one can identify $\tilde\omega_i$ with
the tangent vector in  $i$-th direction. This assignment is invariant: a
linear change of coordinates $H_i$ involves an inverse linear change
in terms of differentials due to the equation of the curve. This
is in agreement with the tensor low for the tangent
space. Hence a tangent vector deforming the spectral curve is identified with
a holomorphic differential $w\in H^0(\K_{\S}).$

\subsection{Riemann bilinear form and dynamics}
Another point of view on the phase space of the
system on the space of rational matrices
involves holomorphic and meromorphic differentials.
As we have already seen the meromorphic differentials
on the spectral curve correspond to
the deformations of the line bundle and the holomorphic differentials
give the deformations of the spectral curve.
Using the splitting of the exact sequence (\ref{ex2}) we can establish an
isomorphism between the tangent space to the phase space
and the meromorphic cohomology space $H^0(\S,\Omega_{\frak{M}}^1),$
where $\Omega_{\frak{M}}^1$ is the space of meromorphic differentials
of the second kind.
There is a natural bilinear form on $H^0(\S,\Omega_{\frak{M}}^1)$
\begin{equation}
\label{Rie}
<\omega_1,\omega_2>=\omega_R(\omega_1,\omega_2)=
\sum_{i=1}^g\left(\int_{a_i}\omega_1\int_{b_i}\omega_2-
\int_{b_i}\omega_1\int_{a_i}\omega_2\right).
\end{equation}
Now we choose the normalized bases of differentials: $\omega_i$ -- holomorphic
differentials and $\Omega_i$~ --~ meromorphic differentials of the second kind
with poles of order $i+1$ at the chosen point~$P_0$ such that
$$\int_{a_j}\omega_i=\delta_{ij};\qquad \int_{a_i}\Omega_j=0.$$
In this basis the matrix of the Riemann bilinear form (\ref{Rie}) is
$$
\omega_R=\left(
\begin{array}[c]{rr}
0 & B\\
-B & 0
\end{array}
\right)
$$
where $B_{ij}=\int_{b_i}\Omega_j.$
{\Th \label{t1} The symplectic structure associated with the canonical
Poisson structure on $\mathcal{T}^*\C^{2g}$  viewed
as a bilinear form on the space of meromorphic differentials coincides
with the Riemann bilinear form $\omega_R.$}
\\
{\bf Proof.} We proceed by evaluating the Hamiltonian map.
As we have seen a flow corresponding
to the Hamiltonian $H_k$ linearizes by
$\int^P\tilde\omega_k$ where $$\tilde\omega_k=\frac {R_k d\l}{\partial_\mu R}.$$
One has a linear decomposition $$\tilde\omega_k=\sum_i A_{ki}\omega_i.$$
A tangent vector $V_k$ on the Jacobian defined as the hamiltonian flow of $H_k$
can be represented as the $k$-th column of $A^{-1}$ in the
basis related to chosen holomorphic differentials $\omega_i.$
Let us note that $dH_k$ has the same coordinates in terms of
$d\tilde H_i,$ where $\tilde H_i$ correspond to $\omega_i.$
The  meromorphic differential
deforming the line bundle $\Omega_0,$ defined by (\ref{mer})
 for the $k$-th flow,
can be given in coordinates associated with the chosen basis of differentials
by $\Omega_0=B^{-1}_{ij}A^{-1}_{jk}\Omega_i.$
It demonstrates that the Hamiltonian vector field related to the
meromorphic differential $\Omega_0$ is obtained by $\beta(dH_k),$ where $\beta$ is the
bivector corresponding to the symplectic form $\omega_R.$ This verifies that
the off-diagonal blocks of comparing symplectic structures coincide. The diagonal
blocks are zero in both cases.
$\blacksquare$

In the dual language this bilinear form can be represented by a
very natural structure. The cotangent space to the phase space of
considered system can be identified with $H_1(\S,\C).$
{\lem \label{ind} The Poisson
bivector is the intersection index form in terms of cycles $H_1(\S,\C).$}
\\
{\bf Proof.} A bivector corresponding to the symplectic
form $\omega$ is defined as follows: the symplectic form defines a
map  $\omega^\#:\mathcal{T}(V)\longrightarrow \mathcal{T}^*(V)$
$$\omega^\#(v)=i(v)\omega.$$
Now a bivector is defined by
$$\beta(\omega^\#(X),\omega^\#(Y))=\omega(X,Y).$$
A covector $\omega^\#(X)$ can be represented by
$$\omega^\#(X)=\sum_i\left(\int_{a_i}X\right)b_i-
\sum_i\left(\int_{b_i}X\right)a_i;$$
one has a similar expression for  $\omega^\#(Y).$
Inserting these covectors into $\beta$ for all $X,Y$ one obtains
$$\beta(b_i,b_j)=\beta(a_i,a_j)=0;\qquad\beta(a_i,b_j)=\delta_{ij};
\qquad \beta(b_i,a_j)=-\delta_{ij}.$$
This bivector coincides with the intersection index.$\blacksquare$
{\Rem The proof of the theorem was given in the case of meromorphic
differentials with one fixed pole of high order. We used this condition
only to normalize differentials. The proof in the general case is strictly analogous.}

\section*{Conclusion}

\begin{itemize}
  \item Lemma \ref{ind} gives us intuition of a geometric interpretation of
the quantization: locally the classical Poisson algebra can be identified with
a free commutative algebra generated by $1$-cycles $H_1(\S,\Z)$ with the
Poisson structure given by the intersection index.
There is an obvious but rather tautological quantization of this Poisson
algebra in geometric terms. One has a noncommutative
multiplication on $H^*(\S)$ which is just a wedge product.
On the first cohomologies it restricts as $$H^1(\S)\times H^1(\S)\longrightarrow
H^2(\S).$$ Due to Poincar\'e duality one can identify $H^1(\S,\Z)$ with
$H_1(\S,\Z)$ and $H^2(\S,\Z)$ with $H_0(\S,\Z).$ In fact this
anticommutative multiplication on cycles coincides with the intersection
index $$c_1*c_2=\#(c_1,c_2).$$ It corresponds to the quantization condition up to
the constant $2$ $$c_1*c_2-c_2*c_1=2\#(c_1,c_2).$$

A question here is to investigate different ways of quantizing this Poisson
algebra and to understand its geometric meaning. Some recent advances in
this direction were realized in \cite{Smi} where the deformed
abelian differentials were introduced and a noncommutative multiplication
over them was constructed. Let us mention that this multiplication is far
from being trivial, however it is anticommutative.
  \item We hope that the appearance of the new object in the context of Hitchin-type
systems -- the lagrangian connection, which we used for splitting the
factorization of the tangent space to the phase space, is not accidental.
Eventually it is not significant on the classical level. However it contains
some intrinsic geometrical data which could be responsible for quantum
deformations.
\end{itemize}

\end{document}